# Visual question answering-based image-finding generation for pulmonary nodules on chest CT from structured annotations

Maiko Nagao[1], Kaito Urata[1], Atsushi Teramoto[1],

Kazuyoshi Imaizumi[2], Masashi Kondo[2], Hiroshi Fujita[3]

[1] Graduate School of Science and Engineering, Meijo University,

[2] Fujita Health University, [3] Gifu University

## Abstract

**[Purpose]** Interpretation of imaging findings based on morphological characteristics is important for diagnosing pulmonary nodules on chest computed tomography (CT) images. In this study, we constructed a visual question answering (VQA) dataset from structured data in an open dataset and investigated an image-finding generation method for chest CT images, with the aim of enabling interactive diagnostic support that presents findings based on questions that reflect physicians' interests rather than fixed descriptions.

**[Methods]** In this study, chest CT images included in the Lung Image Database Consortium and Image Database Resource Initiative (LIDC-IDRI) datasets were used. Regions of interest surrounding the pulmonary nodules were extracted from these images, and image findings and questions were defined based on morphological characteristics recorded in the database. A dataset comprising pairs of cropped images, corresponding questions, and image findings was constructed, and the VQA model was fine-tuned on it. Language evaluation metrics such as BLEU were used to evaluate the generated image findings.

**[Results]** The VQA dataset constructed using the proposed method contained image findings with natural expressions as radiological descriptions. In addition, the generated image findings showed a high CIDEr score of 3.896, and a high agreement with the reference findings was obtained through evaluation based on morphological characteristics.

**[Conclusion]** We constructed a VQA dataset for chest CT images using structured information on the morphological characteristics from the LIDC-IDRI dataset. Methods for generating image findings in response to these questions have also been investigated. Based on the generated results and evaluation metric scores, the proposed method was effective as an interactive diagnostic support system that can present image findings according to physicians' interests.

## 1. Introduction

Cancer is one of the leading causes of death worldwide, and lung cancer has high incidence and mortality rates [1]. Low-dose computed tomography (CT) screening is effective in reducing lung cancer mortality [2], and the importance of evaluating pulmonary nodules based on chest CT images is expected to increase in the future. However, differentiating benign from malignant pulmonary nodules requires consideration of multiple factors, such as shape, internal characteristics, and relationships with surrounding tissues, and the diagnostic process still largely depends on radiologists' experience.

In recent years, automatic classification methods using deep learning have improved the accuracy of distinguishing between benign and malignant pulmonary nodules [3]. However, because many existing methods provide only classification results, interpretability issues remain. To address this limitation, we previously proposed a method that uses vision-language models (VLMs) to generate image findings from chest CT images and simultaneously perform benign-malignant classification [4]. However, this method assumes that image generation is fixed and does not sufficiently support the interactive diagnostic process in clinical practice.

In image-based diagnosis of pulmonary nodules, comprehensive decisions are made by sequentially evaluating the morphological characteristics. Visual question answering (VQA), which can present image findings based on images and questions, has attracted attention as a method for supporting such diagnostic processes. Although several studies on VQA for medical imaging have been reported in recent years [5], they have not adequately addressed the generation of image findings that reflect clinically important information for pulmonary nodule diagnosis, such as shape and internal characteristics, nor have they supported a systematic evaluation based on morphological features. In addition, datasets suitable for training medical VQA models that include high-quality correspondences among images, questions, and answers remain limited.

In this study, we focused on the lung image database consortium and image database resource initiative (LIDC-IDRI) dataset, which contains structured annotations of the morphological characteristics. We constructed a VQA dataset comprising chest CT images, questions about morphological characteristics, and the corresponding image findings. Furthermore, using the constructed dataset, we implemented a VQA framework that enabled the interactive generation of image findings from chest CT images and investigated its effectiveness.

## 2. Materials and Methods
### 2.1 Overview

An overview of the proposed method is shown in Fig. 1. Pulmonary nodule regions were extracted from chest CT images in the LIDC-IDRI dataset, and a VQA dataset was constructed using the structured data recorded in the database. The VQA model was fine-tuned using the VQA dataset. The model's ability to generate responses related to the nodule image findings and to describe the morphological characteristics was then evaluated.

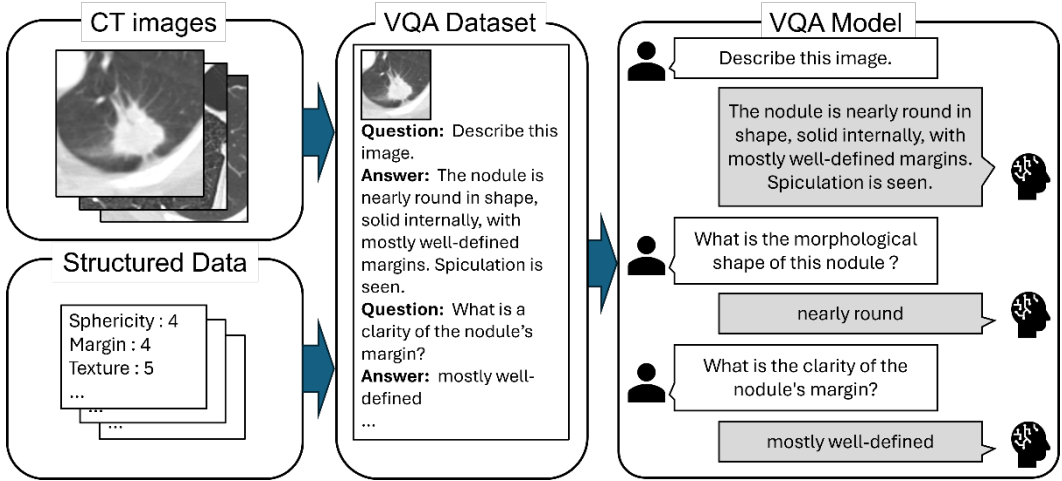

Fig. 1. Schematic overview of the proposed framework.

## 2.2 VQA dataset construction

In this study, we introduced the LIDC-IDRI dataset, which includes chest CT images from 1018 cases with detailed annotations from up to four radiologists. These annotations record structured data describing the morphological characteristics of the pulmonary nodules. For each nodule, a square region of interest (ROI) with a side length equal to twice the long-axis diameter of the nodule was defined with the nodule center as the center of the ROI, and image cropping was performed. CT images were obtained from DICOM files, and the single slice with the z-coordinate closest to the nodule center was selected.

Based on the morphological characteristics described in the structured data, six features were considered: sphericity, margin, texture, lobulation, spiculation, and calcification. These features were scored on a scale of 1–5 (1–6 for calcification). As the LIDC-IDRI dataset provides interpretation results from four radiologists for each nodule, the median of the radiologists' scores was calculated and rounded to the nearest integer to define a single representative value. For the obtained morphological feature scores, the corresponding linguistic expressions were predefined for each score value, and the image findings for each nodule were generated using these expressions. An example of the constructed dataset is shown in Fig. 2.

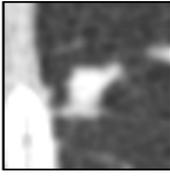

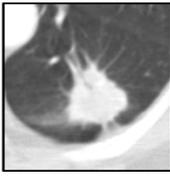

Fig. 2. Examples of the VQA dataset.

To generate image findings for pulmonary nodules, in addition to a question querying the overall image findings, questions corresponding to the six morphological characteristics recorded in the XML files were defined, and seven types of questions were used as inputs. Table 1 presents specific examples of the questions used in this study.

Table 1. Examples of question prompts

| Category | Target finding | Question prompt |
|---|---|---|
| Overall finding | Image finding | Describe this image. |
| Morphology | Sphericity | What is the morphological shape of this nodule? |
| | Margin | What is the clarity of the nodule's margin? |
| | Texture | What is the internal structure of this nodule? |
| | Spiculation | Does this nodule exhibit spiculation? |
| | Lobulation | Does this nodule exhibit lobulation? |
| | Calcification | What is the type of calcification present in this nodule? |

## 2.3 VQA model

In this study, bootstrapping language-image pre-training (BLIP) was used as the VQA model to integrate visual and linguistic information. BLIP is a multimodal model developed by Salesforce that utilizes both images and natural language [6]. It is pretrained on large-scale image–text pairs, extracts visual features using a vision transformer [7], and performs natural language generation using bidirectional encoder representations from transformers (BERT) [8], which is a natural language processing model. The architecture of the BLIP is illustrated in Fig. 3.

As the purpose of this study was to verify the effectiveness of the constructed dataset rather than to compare model performance, BLIP was adopted because it has a relatively small parameter count and enables stable training with limited data.

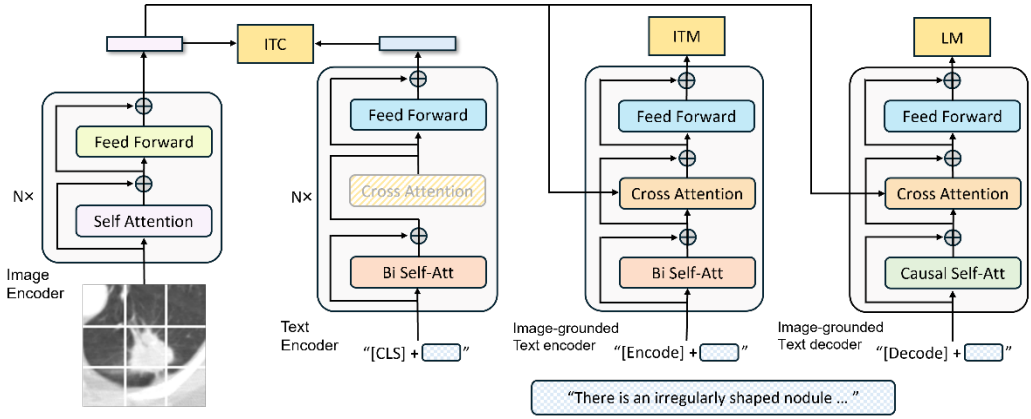

Fig. 3. Structure of BLIP.

### 2.4 Training and evaluation strategy

A pre-trained BLIP model was fine-tuned using a constructed dataset of nodule images and their corresponding findings. The Adam optimizer was used with a learning rate of $1 \times 10^{-5}$ and a batch size of eight. The maximum number of training epochs was set as 20. At the end of each epoch, the CIDEr score [9] was computed on the validation dataset. The model with the highest CIDEr score was selected as the final one. The dataset was split into training, validation, and test sets in a 7:2:1 ratio, with 1453, 416, and 208 images, respectively. All the training and inference experiments were conducted on a PC with an AMD Ryzen 9 5900X CPU and an NVIDIA GeForce RTX 3090 GPU.

For quantitative evaluation of the generated image findings, five metrics were used: BLEU, METEOR, ROUGE-L, CIDEr, and SPICE [10–13]. Additionally, for clinical evaluation, we assessed the agreement between the three morphological feature scores extracted from the generated image and the reference findings. The agreement was assessed using the mean absolute error (MAE) and consistency, which were calculated using Equations (2.1) and (2.2).

In these equations, $N$ denotes the number of nodules included in the evaluation, $\hat{y}_i$ represents the score of the morphological characteristics of the reference findings, and $y_i$ represents the score derived from the generated image findings. $D_{max}$ denotes the difference between the maximum and minimum possible scores for each morphological characteristic. For the three morphological characteristics evaluated in this study, $D_{max} = 4$.

$$MAE = \frac{1}{N}\sum_{i=1}^{N}|\hat{y_i} - y_i| \qquad (2.1)$$

$$consistency = 1 - \frac{1}{N \cdot D_{max}}\sum_{i=1}^{N}|\hat{y_i} - y_i| \qquad (2.2)$$

## 3. Results and Discussion

Examples of the generated image findings are presented in Fig. 4, and the evaluation metrics for image-finding generation are summarized in Table 2. Table 3 lists the MAE and consistency results for each morphological characteristic.

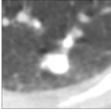

Fig. 4. Example of output. (a) A correctly generated finding consistent with the reference annotation. (b) An example showing discrepancies between the generated finding and the reference annotation.

Table 2. Evaluation metrics

| BLEU_1 | BLEU_2 | BLEU_3 | BLEU_4 | METEOR | ROUGE_L | CIDEr | SPICE |
|---|---|---|---|---|---|---|---|
| 0.869 | 0.801 | 0.727 | 0.662 | 0.493 | 0.875 | 3.896 | 0.862 |

Table 3. Quantitative evaluation of morphological feature prediction

|  | Sphericity | Margin | Texture |
|---|---|---|---|
| MAE | 0.668 | 0.716 | 0.418 |
| Consistency | 0.833 | 0.821 | 0.895 |

As shown in Table 2, the generated image findings achieved a BLEU-4 score of 0.662 and CIDEr score of 3.896, demonstrating a high level of correspondence with the reference findings in terms of both linguistic quality and content. In a previous study, a BLEU-4 score of 0.561 and a CIDEr score of 0.903 were reported for a BLIP trained on a single-institution dataset. The results obtained in the present study exceeded these values, suggesting the effectiveness of training with a large-scale dataset that includes structured information on morphological characteristics.

A previous study that used machine learning to score morphological characteristics reported MAE values of 0.83 ± 0.75, 0.83 ± 0.86, and 0.46 ± 0.84 for sphericity, margin, and texture, respectively [14]. Comparing these values and those in Table 3, the MAE values for all morphological characteristics in this study were better. Furthermore, because the consistency values exceeded 0.8 for all characteristics, the proposed method accurately evaluated morphological characteristics.

However, the questions and morphological characteristics addressed in this study are limited. In future studies, expanding the types of questions and target image-finding items will enable more clinically relevant interactive diagnostic support. Additionally, although image-finding generation in this study was performed using pre-extracted nodule regions, further investigations are required to generate findings from whole-chest CT images and integrate the proposed method with nodule detection.

## 4. Conclusion

In this study, a VQA dataset comprising chest CT images, questions about morphological characteristics, and the corresponding imaging findings was constructed from the LIDC-IDRI dataset. Furthermore, interactive generation of image findings from chest CT images was enabled. Based on the generated image findings and the values of the evaluation metrics, the proposed method is potentially effective.

**Acknowledgements**

This study was supported in part by a Grant-in-Aid for Scientific Research (No. 23K07117)

from the Ministry of Education, Culture, Sports, Science and Technology, Japan.

**Conflict of Interest**

The authors declare that they have no conflict of interest.

**Ethical approval**